\documentclass[a4paper,12pt, nofootinbib]{revtex4}

\usepackage{graphicx}
\usepackage{bbm}

\begin{document}

\title{Gravity~in~quantum~spacetime\footnote{\baselineskip 14pt This essay received honorable mention in the Gravity Research Foundation 2010 Awards for Essays on Gravitation}}

\author{Giovanni AMELINO-CAMELIA}
\affiliation{Dipartimento di Fisica, Universit\`a di Roma ``La Sapienza''\\ and Sez.~Roma1 INFN, Piazzale Aldo Moro 2, 00185 Roma, Italy}
\email{Giovanni.Amelino-Camelia@roma1.infn.it}
\author{Niccol\`{o} LORET}
\affiliation{Dipartimento di Fisica, Universit\`a di Roma ``La Sapienza''\\ and Sez.~Roma1 INFN, Piazzale Aldo Moro 2, 00185 Roma, Italy}
\author{Gianluca MANDANICI}
\affiliation{Dipartimento di Fisica, Universit\`a di Roma ``La Sapienza''\\ and Sez.~Roma1 INFN, Piazzale Aldo Moro 2, 00185 Roma, Italy}
\author{Flavio MERCATI}
\affiliation{Dipartimento di Fisica, Universit\`a di Roma ``La Sapienza''\\ and Sez.~Roma1 INFN, Piazzale Aldo Moro 2, 00185 Roma, Italy}

\begin{abstract}
The literature on quantum-gravity-inspired scenarios for the quantization
of spacetime has so far focused on particle-physics-like studies.
This is partly justified by the present limitations of our understanding of
quantum-gravity theories, but we here argue that
valuable insight can be gained through semi-heuristic analyses
of the implications for gravitational phenomena of some results
obtained in the quantum-spacetime literature.
In particular, we show that the types of description
of particle propagation that emerged in certain quantum-spacetime frameworks
have striking implications for gravitational collapse and for the behaviour
of gravity at large distances.
\end{abstract}

\maketitle

\baselineskip 14pt

\newpage

\section{Introduction}

In quantum-gravity research it is natural to contemplate
the possibility of (one form or another of) spacetime
quantization. The vigorous research effort devoted
to this hypothesis
has been focused on cases where the quantum-spacetime
effects could be studied neglecting the ``gravity aspects"
of the quantum-gravity problem.
We here argue that this strategy,
partly justified by our present technical limitations
in the analysis of the hugely complex candidates for
quantum gravity,
might produce logically inconsistent results or at least
lead us to an incorrect intuition for the quantum-gravity realm.

We provide support for our thesis through a semi-heuristic/semi-quantitative
investigation of the implications of spacetime quantization
for certain significant classes of gravitational phenomena.
And specifically
we illustrate
our concerns by considering the laws of propagation of particles in certain
much-studied examples of quantum spacetime,
which have been investigated in the relevant
literature~\cite{majidruegg,gampul,gacmajid,urrutiaPRLePRD,gacNATUREqgphen,kowaTIMETORIGHT,szaboREVIEW}
mainly through the
 lens of modifications of the energy-momentum
dispersion relation $E^2 = m^2 + p^2 + \Delta_{QST}(p;L_P)$.
Different quantum-spacetime frameworks
lead to different results and intuition for
the dependence of the quantum-spacetime correction
term $\Delta_{QST}$ on the particle wavelength $1/p$
and on the ``Planck length" $L_P (\sim 10^{-35} m)$.
Of particular significance from our perspective is the
difference between the (more frequent)
examples of $\Delta_{QST}$
that exclusively affect the short-wavelength/ultraviolet (UV) regime
and the few examples of quantum-spacetime pictures in
which $\Delta_{QST}$ is actually significant in the long-wavelength/infrared (IR) regime.

We here give two illustrative examples of the strategy of analysis
that we are advocating. For the case of long-wavelength effects we consider some
possible implications for the behaviour of gravity at large distances.
And for the case of short-wavelength effects we consider some possible
implications for gravitational collapse.

\section{Modified Netwonian gravity at large distances}

We start with an observation that concerns the possibility of
IR modifications of the dispersion relation.
While most of quantum-gravity research has focused on the traditional ``short-distance
paradigm" of the last century of fundamental physics, it is actually rather natural
to expect that quantum gravity should also have profound implications in the infrared.
This was perhaps most eloquently advocated in Ref.~\cite{cohenUVIR},
on the basis of
our present understanding
of (quantum-) black-hole thermodynamics,
and finds some
support in parts of the quantum-gravity/quantum-spacetime literature.
Of particular interest for our purposes is the case of modifications
of the dispersion relation which in the long-wavelength regime take the form\footnote{We
adopt ``natural units" $\hbar = c =1$.}
\begin{equation}
E^2 \simeq m^2 + p^2 - \xi p~.
\label{mdrurrutia}
\end{equation}
These were found in a quantum-spacetime model inspired by one of the competing perspectives
on the  (still unknown)
semiclassical limit of Loop Quantum Gravity~\cite{urrutiaPRLePRD} (also see Ref.~ \cite{josePLBePRD})
 and is also the
qualitative behaviour\footnote{\baselineskip 14pt Light-like
noncommutativity with UV supersymmetry produces~\cite{szaboREVIEW}
corrections to the dispersion relation with IR behaviour
of type $\log (1+p_*/m)\propto p_*$ (where $p_*$ is the
spatial momentum in a preferential direction determined by the noncommutativity matrix).}
of the dispersion relation in the IR regime
found~\cite{szaboREVIEW} in the study of spacetime noncommutativity in the so-called ``light-like
noncommutativity models" (assuming UV supersymmetry~\cite{szaboREVIEW}).
Of course, one does not expect the energy scale $\xi$, introduced to characterize
the long-wavelength regime,
 to be ``Planckian", and indeed the relevant models encourage the expectation that $\xi \ll 1/L_P$.
The scale of onset of IR effects for light-like noncommutativity can be
described~\cite{szaboREVIEW}
 as a ratio
formed with an energy scale $M_{NC}$ that characterizes the noncommutativity matrix $\theta_{\mu \nu}$
($\theta_{\mu \nu} = - i [x_\mu ,  x_\nu]$)
and a much higher cutoff energy scale $\Lambda$, suggesting $\xi \sim M^2_{NC}/\Lambda$.
A somewhat similar mechanism favours small values of the energy scale $\xi$
also within the relevant Loop-Quantum-Gravity-inspired
model~\cite{urrutiaPRLePRD,gacFlavioPRL2009}.

We here point out that
a noteworthy gravitational implication of (\ref{mdrurrutia})
concerns the regime where gravity is governed by the Newtonian potential.
The Newtonian potential is produced by a static point source when the field that mediates the
force described by the potential has energy-momentum space (inverse) propagator $G^{-1}(E,p)=E^2-p^2$.
In general, if the field that mediates the force has different propagator,
as in the case of our $G_\xi^{-1}(E,p) \simeq E^2-p^2+\xi p$,
the Newtonian potential produced at the spatial point $\vec r$ by a point-like mass $M$, located
 at the origin,
is replaced by the potential obtained by computing~\cite{peskinBOOK,hellingCOULOMBscreen}
\begin{equation}
V(\vec r) = L_p^2 M \int \frac{d^3p}{2 \pi^2}~G(0,\vec p) ~e^{i \vec p \cdot \vec r} ~,
\label{PotenzialeDaPropagatore}
\end{equation}
{\it i.e.} the potential is the spatial Fourier transform
of the propagator evaluated at $E = 0$. In the case here of interest we must therefore
consider the $\xi$-deformed Newtonian potential
$$
\frac{V_\xi(r)}{ L_p^2 M}= \int \frac{d^3p}{2 \pi^2}~\frac{e^{i \vec p \cdot \vec r}}{G_\xi^{-1}(0,p)}
= \frac{1}{r}\frac{2}{\pi}\int_0^\infty\frac{\sin(p r)}{p}\frac{p^2}{p^2-\xi p}dp=\frac{1}{r}\frac{2}{\pi}\int_0^\infty\frac{\sin(p r)}{p-\xi}dp~.
$$
We are working essentially from the viewpoint of some sort
of ``high-energy effective theory"~\cite{cohenUVIR},
probing infrared features as we approach them from the regime of higher momenta
(just like in the more standard low-energy effective theories we investigate the ultraviolet
regime by approaching it from below). Accordingly we are interested in a small-$r$
expansion, and for these purposes it is useful to change variable of
integration $p \rightarrow p-\xi \equiv q$ and make the following observations:
\begin{eqnarray}
\frac{V_\xi(r)}{ L_p^2 M}&=&\frac{1}{r}\frac{2}{\pi}\int_{-\xi}^\infty\frac{\sin((q+\xi) r)}{q}dq =\nonumber\\
&=& \frac{1}{r}\frac{2}{\pi}\left(\cos(\xi r)\int_{-\xi}^\infty\frac{\sin(q r)}{q}dq
 +\sin(\xi r)\int_{-\xi}^\infty\frac{\cos(q r)}{q}dq\right)=\nonumber\\
&=&\frac{1}{r}\frac{2}{\pi}\left(\cos(\xi r)\left(\int_0^\infty\frac{\sin(q r)}{q}dq+\int_0^{r\xi}\frac{\sin(t)}{t}dt\right)+\sin(\xi r)\int_{\xi r}^\infty\frac{\cos(t)}{t}dt\right)=\nonumber\\
&=&\frac{1}{r}\frac{2}{\pi}\left(\cos(\xi r)\left(\frac{\pi}{2}+{\mathrm{Si}}(\xi r)\right)-\sin(\xi r){\mathrm{Ci}}(\xi r)\right)\simeq\frac{1}{r}-\frac{2}{\pi}\xi\ln(r) ~. \label{logr}
\end{eqnarray}

As shown in Fig.~1 this result (\ref{logr}) for $V_\xi(r)$
could have significant implications for $r$ of the order of (but smaller than) $\xi^{-1}$:
in that regime (\ref{logr}) produces a $\xi$-modified gravitational force
with behavior $\simeq 1/r^2 + (2/\pi) \xi/r$.
As also suggested by Fig.~1 (dashed-thin line)
we are proposing to analyze this type of issues from a ``high-energy effective theory" perspective
also for what concerns the possibility of subleading IR terms: for example adding to our
$G_\xi^{-1}(E,p)$ terms of the type $\xi^3/p$ produces sizeable (and, from the perspective of effective
theory, largely uncontrollable) modifications as $r$ gets close to $\xi^{-1}$ from below,
but can be safely neglected for smaller values of $r$.

It is amusing to notice that a strengthening of gravity at large distances,
and indeed one with a qualitative behavior roughly of type $1/r$,
has been advocated~\cite{mondREVIEW} to describe the much-debated unexplained features
of galaxy rotation curves.
We shall explore elsewhere~\cite{noiINPREP} to what extent
a reasonably realistic MOND-like~\cite{mondREVIEW} scenario
can be produced in our framework. We do notice here that a conceptual
perspective inviting such considerations is offered
by Ref.~\cite{gacSMOLIN2004}, where it was independently argued
that in non-flat spacetimes the Planck-scale modifications
of the laws of particle propagation should be ``context dependent", in the
sense that their magnitude should depend on the local values of some geometric observables
(such as the Ricci scalar considered in  Ref.~\cite{gacSMOLIN2004}).

\begin{figure}[h!]
\begin{center}
\includegraphics[width=0.48\textwidth]{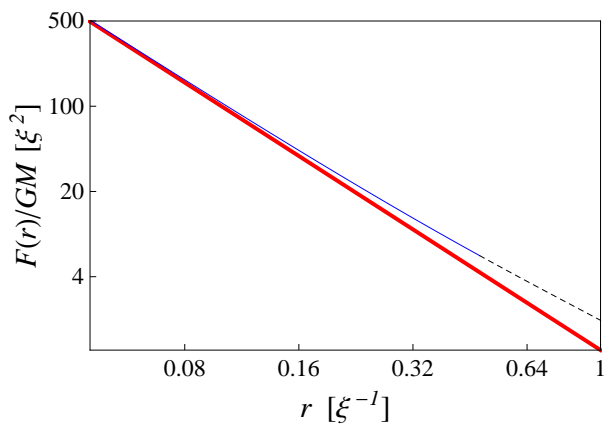}
\caption{\baselineskip 14pt We here characterize graphically (in a ``log/log plot")
the modifications of gravity in the Newtonian regime induced
by the parameter $\xi$. We compare the standard ($1/r^2$)
Newtonian force (thick line) to the $\xi$-modified Newtonian force (thin line).
Of interest to us is the modification of Newtonian gravity
found as $r$ approaches the value $\xi^{-1}$ from below, {\it i.e.} approaching the
infrared/long-distance regime in the sense of effective theory.
 And there we find that
the $\xi$-corrected force has behavior that differs from the standard $r^{-2}$ law
by a term which roughly behaves like $r^{-1}$. For values of $r$ that get very close to $\xi^{-1}$
we (unsurprisingly) find that the prediction becomes very sensitive to possible subleading
IR corrections (and we give visible warning for this issue by using a dashed-thin line
for the modified Newtonian force for those values of $r$).}
\end{center}
\end{figure}

\section{Fermi pressure and gravitational collapse}

Our second illustrative example of the possible implications for gravitational
phenomena of some quantum-spacetime effects focuses instead on
 the  case of ``UV modifications"
of the dispersion relation,
such that in leading order in the Planck length
\begin{equation}
E^2 \simeq m^2 + p^2 + \eta L_P^n p^{2+n}~,
\label{ModDispRelLeading}
\end{equation}
where the model-dependent integer $n$ ($\geq 1$)
fixes the dependence on wavelength (and the Planck length)
of the leading-order correction, and $\eta$ is a parameter expected to be roughly of order 1.
This type of modifications of the dispersion relation are found in the analysis
of certain models of ``Lie-algebra-type spacetime
noncommutativity"~\cite{majidruegg,gacmajid,kowaTIMETORIGHT}
and have also been advocated~\cite{gampul,thiemLQG,leeLQGDSR1} within some
perspectives on the
semiclassical limit of ``Loop Quantum Gravity".
The intuition that drives these proposals is that in the quantum-gravity realm
spacetime cannot be described as a smooth classical geometry, but rather it should have
rich (and dynamical) short-distance structure, affecting in different ways particles
of different wavelength (long-wavelength particles could be essentially unaffected because
of a coarse-graining mechanism).

\noindent
These UV modifications of the dispersion relation have been studied
extensively from a particle-physics perspective~\cite{majidruegg,gampul,gacmajid,urrutiaPRLePRD,gacNATUREqgphen,kowaTIMETORIGHT,szaboREVIEW},
but we are interested in their relevance
for the description of gravitational collapse, which we shall here
be satisfied to investigate preliminarily, essentially
in the framework of the Landau model~\cite{landauNEUTRONSTAR}.

The core mechanism of gravitational collapse concerns the fact that when the total mass of a gas
of self-gravitating fermions (neutrons)
is increased above a certain critical value the gravitational
pressure cannot be balanced by
the ``Fermi pressure" (the degeneracy pressure resulting from the Pauli exclusion principle).
In a classical-spacetime picture
the gravitational collapse advances all the way to the production of a singularity, {\it i.e.}
the case of the whole mass of the system collapsed to a point.
We find that the UV effects described in Eq.~(\ref{ModDispRelLeading})
can modify this picture, and actually there are two features that act in the same
direction of producing the possibility of equilibrium between Fermi pressure and gravitational
pressure even for masses above the critical value.
One of these features concerns a weakening of gravitational pressure
deduced by assuming
the $L_P$-modified dispersion relation (\ref{ModDispRelLeading})
in the evaluation of the gravitational potential (\ref{PotenzialeDaPropagatore}).
The other feature concerns a modification
of the Fermi pressure:
the $L_P$-modified dispersion relation (\ref{ModDispRelLeading}) is such that more energy is needed
(with respect to the $L_P \rightarrow 0$ classical-spacetime case) in order to ``squeeze" the system
by pushing more fermions to states with ultrashort wavelength.
As a result we find that
for the case $n=1$ the dependence of the ``speed of sound"\cite{Zeldovich} on the radius
of the system $R$ is
\begin{equation}
v_s^2 = \frac{\partial P}{\partial \epsilon} \simeq \frac{1}{3} + \frac{\eta}{3} (3\pi)^{2/3} L_P \frac{N^{1/3}}{R}
\label{sound}
\end{equation}
This means that, unlike the classical-spacetime case,
in these quantum-spacetime pictures one has violations of the  $v_s^2 \leq 1/3$ constraint\cite{Zeldovich}.
And it is noteworthy that
the Planck-length correction is ``amplified" by the number $N$  of fermions in the system.
This is due to the Pauli exclusion principle, which implies that some fermions have Planckian momentum
(the Fermi momentum $p_F$ is $\sim 1/L_P$) even when the average momentum in the system is
below the Planck scale.

As shown in Figure~2 our Planck-scale correction modifies Newtonian gravity, by softening it,
with effects that are however only significant when the size of the system is of the
order of the Planck length. Instead, as also shown in Figure~2, the mentioned amplification
of the Planck-scale effects by the number of fermions
produces a significant increase of the speed of sound (and of the Fermi pressure)
 already for relatively macroscopical sizes
of the system, depending on both the Planck length and the total mass of the gas of neutrons.
Both effects go in the direction of producing equilibrium configurations even for masses
that are above the critical value, but the modification of Fermi pressure kicks in first
(when the size of the system is still much larger than Planckian), so that, at least within
our simplified analysis, it dominates over the effect of modification of the gravitational
pressure. This is illustrated in Figure~3.
\begin{figure}[h!]
\begin{center}
\includegraphics[width=0.48\textwidth]{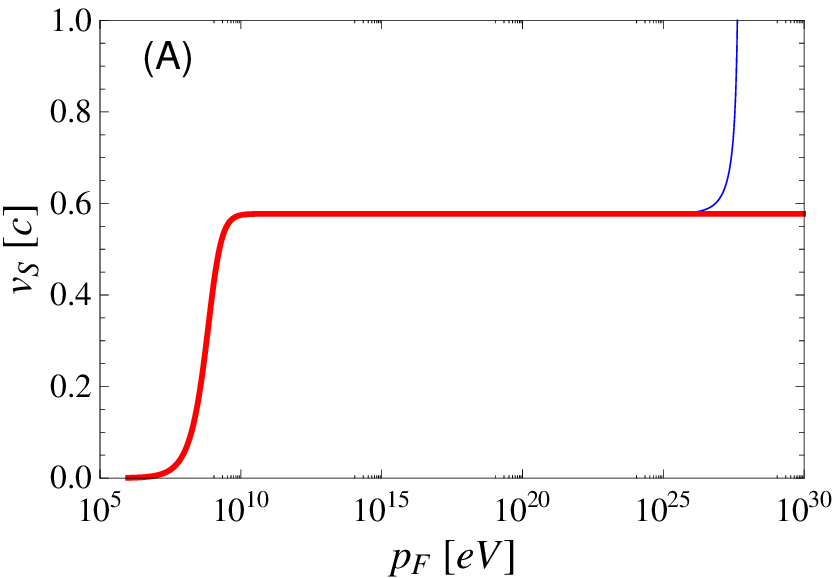}
\includegraphics[width=0.48\textwidth]{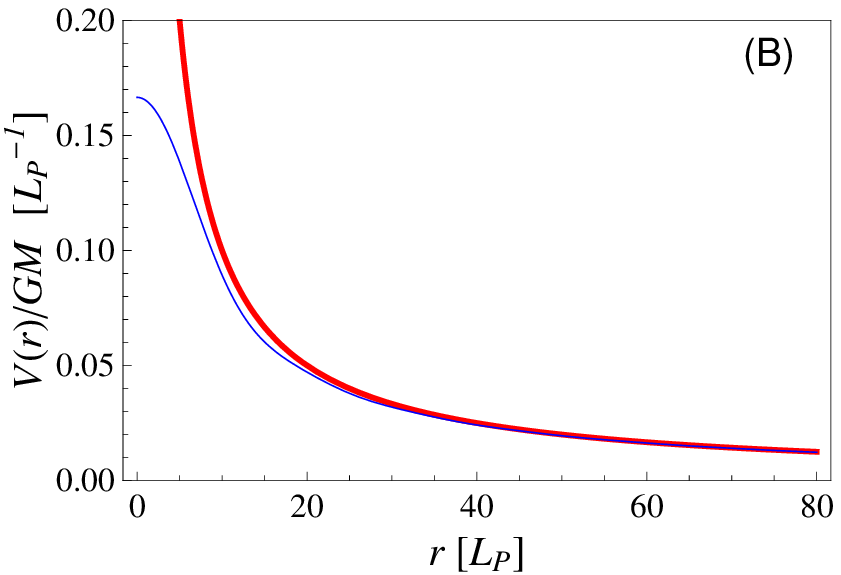}
\caption{\baselineskip 14pt We here characterize graphically the behaviour
of the $L_P$-modified speed of sound (panel A) and  Newtonian potential (panel B)
found for the case of the dispersion relation given in Eq.~(\ref{mdrDSR1}).
Thick lines are for the standard case while thin lines are for the case of Eq.~(\ref{mdrDSR1}).
It is noteworthy that the UV correction to the speed of sound becomes significant when the Fermi momentum
is ``Planckian", which occurs when the size of the system is still much larger than the Planck length.
The UV correction to the Newtonian potential is instead only significant when the size
of the system is of the order of the Planck length.
}
\end{center}
\end{figure}

\begin{figure}
\begin{center}
\includegraphics[width=0.48\textwidth]{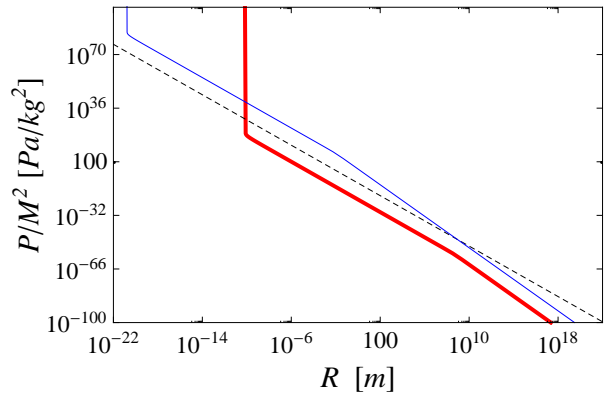}
\caption{\baselineskip 14pt We here assume Eq.~(\ref{mdrDSR1}) and
compare  (in a ``log/log plot") the Fermi pressure
and the gravitational pressure, both divided by the square of the total mass of the system.
A configuration of equilibrium
is given by a point where the Fermi pressure equals
the gravitational pressure.
For the gravitational pressure (dashed line) the UV modification
 is not significant (and not visible)
in the range of values of the system radius $R$ that is shown in the figure
(it becomes significant only for $R \simeq L_P$). Two cases are shown for the Fermi pressure,
one assuming total mass of the system much higher than the critical mass (thick line)
and one with mass much smaller than the critical value (thin line).
}
\end{center}
\end{figure}

\noindent
Both Figure~2 and Figure~3 are drawn specifically for the case $n=1$, $\eta = 2$,
and actually assuming
\begin{equation}
 \sinh^2 \left( L_P \, E \right) = \sinh^2 \left( L_P \, m\right) + L_P^2 e^{L_P E} p^2 ~,
\label{mdrDSR1}
\end{equation}
which indeed for $p < 1/L_P$ reproduces (\ref{ModDispRelLeading})
for $n=1$, $\eta = 2$.
For the emergence of new equilibrium configurations the leading-order formulas
are already reliable~\cite{noiINPREP}, but it is interesting to contemplate in particular
this candidate for the exact form of the dispersion relation. Eq.~(\ref{mdrDSR1})
 emerges in two different (but related)
formalisms~\cite{majidruegg,gacdsr1}, and
has also generated strong
interest
because it fits the popular intuition that the Planck
length should set the minimum allowed value of wavelength:
according to (\ref{mdrDSR1}) one must confine $p$
to $p \lesssim 1/L_P$, because the energy $E$ diverges already as $p$ approaches from below $1/L_P$.
This provides a noteworthy example of the mechanism discussed above,
since it implies that the gravitational collapse  should necessarily stop
at values of the radius such that $p_F \simeq 1/L_P$, where the system becomes
incompressible (the speed of sound diverges).

\section{Many challenges and a few opportunities}

Even within the confines of the semi-heuristic level of analysis we adopted,
it is noteworthy that
the two illustrative examples on which we centered this first exploratory
analysis based on our novel approach
appear to be relevant for some of the most hotly debated issues
in contemporary fundamental physics. ``Singularity avoidance" is a recurring theme of
quantum-gravity research, and yet to our knowledge the possible relevance of
quantum-spacetime results on UV modifications of the dispersion relation,
established by our Fermi-pressure argument for gravitational-collapse analysis, had been
so far overlooked.
And other issues that are much under the limelight
concern the behaviour of gravity at ultra-large distances, where
the galaxy rotation problem and dark matter must be dealt with. In this respect it was
amusing for us to notice that the
gravitational force could be strengthened at large distances, as a result of IR
modifications of the dispersion relation which had been previously proposed
in the quantum-spacetime literature.

These candidate quantum-spacetime-induced modifications of gravitational phenomena
might be legitimately perceived
from some perspectives as rather exciting opportunities for discovery,
but perhaps we should stress even more strongly the alternative perspective that sees them
as challenges.
There is truly some room for novelty both at ultra-short and ultra-large distance scales,
but this should not obscure the fact that
the present description of gravity is very successful experimentally
in a vast range of distance scales
and in an impressive number of applications.
A realistic goal for our proposed research programme
 is to establish that some quantum-spacetime pictures affect gravity
at levels that contradict known experimental facts, since then, with some prudence only due to
the semi-heuristic strategy of analysis we are advocating, theory work on those quantum-spacetime
proposals could be disfavoured. In the difficult search for quantum gravity even observations
that provide tentative guidance for the directions (not) to be followed on the theory side are rare
and correspondingly valuable.


\end{document}